\documentclass[runningheads]{llncs}
\usepackage{amssymb}
\usepackage{amsmath}
\usepackage{hyperref}
\usepackage{tabularx}
\usepackage{booktabs}
\usepackage{arydshln}
\usepackage{subfigure}
\usepackage{trimclip}
\usepackage{tikz, pgfplots}
\usepackage{ifthen}
\usepackage{mathtools}
\usepackage{listings}
\usepackage{xspace}

\definecolor{seagreen}{rgb}{0.0, 0.26, 0.15}

\usepackage{thm-restate}
\usepackage{refcount}

\newcommand{\new}[1]{\color{black}#1\color{black}\xspace}

\newcounter{reftmpcounter}
\newcounter{resttmpcounter}
\newcommand{\restatethm}[2]{
    \setcounterref{reftmpcounter}{#2}
    \setcounter{resttmpcounter}{\thetheorem}
    \setcounter{theorem}{\thereftmpcounter}
    \addtocounter{theorem}{-1}
    #1
    \setcounter{theorem}{\theresttmpcounter}
}

\newboolean{fullgraphics} 
\setboolean{fullgraphics}{true}

\usepackage{graphicx}
\begin{document}
\newcommand{\w}{w}
\newcommand{\rlow}{q}
\newcommand{\Aocc}{A_{\blacksquare}}
\newcommand{\Aarea}{A_{\square}}
\newcommand{\dlow}{\delta}
\newcommand{\dmin}{\delta_{min}}
\newcommand{\area}{area}
\newcommand{\occ}{occ}
\newcommand{\len}{\ell}
\newcommand{\pl}{p}
\newcommand{\plt}{p_t}
\newcommand{\plb}{p_b}
\newcommand{\fl}{f}
\newcommand{\flt}{f_t}
\newcommand{\flb}{f_b}
\newcommand{\wid}{w}
\newcommand{\den}{den}
\newcommand{\bmdelta}{\delta}
\newcommand{\TP}{\textit{TLP}}
\newcommand{\SP}{\textit{SLP}}
\newcommand{\SqP}{\textit{SP}}
\newcommand{\SPL}{\textit{SPL}}
\newcommand{\EP}{\textit{EP}}
\newcommand{\DSP}{\textit{DSLP}}
\newcommand{\bc}{\color{blue}(}
\newcommand{\ec}{\color{black}) }
\newcommand{\bt}{\color{blue}}
\newcommand{\et}{\color{black}}
\newcommand{\dTP}{d_{TLP}}
\newcommand{\dSP}{d_{SLP}}
\newcommand{\customMin}{\min}
\newcommand{\customMax}{\max}
\newcommand{\pos}{pos}
\newcommand{\extra}{extra}
\newcommand{\qS}{q_S}
\newcommand{\qL}{q_L}
\newcommand{\qD}{q_S}
\newcommand{\qM}{q_M}
\newcommand{\qMin}{q_{min}}
\newcommand{\dstar}{\hat{\delta}}
\newcommand{\qstar}{\qS^*}
\newcommand{\R}{\mathcal{R}}
\newcommand{\E}{\mathcal{E}}
\newcommand{\OES}{\mathcal{O}}
\newcommand{\M}{{min_{SLP}}}
\newcommand{\MDSP}{min_{DSLP}}
\newcommand{\HC}{\mathcal{H}}
\newcommand{\C}{\mathcal{C}}
\newcommand{\f}{f}
\newcommand{\rmin}{r}
\newcommand{\resultOne}{0.350389}
\newcommand{\resultTwo}{0.375898}
\newcommand{\resultThree}{0.369489}
\newcommand{\radiusTwo}{ 0.026622}
\newcommand{\radiusThree}{0.024447}
\newcommand{\worstAreaSquare}{0.396690}
\newcommand{\myl}{\ell}

\title{Online Circle Packing}
%
%
\author{S\'{a}ndor P. Fekete\inst{1}\orcidID{0000-0002-9062-4241}\and \\
Sven von H\"{o}veling\inst{2,}\orcidID{0000-0003-3937-2429}\and \\  
Christian Scheffer\inst{1}\orcidID{0000-0002-3471-2706}}
\authorrunning{S. P. Fekete, C. Scheffer, S. von Hoeveling}
%
\institute{Department of Computer Science, TU Braunschweig, Germany\\ \email{\{s.fekete,c.scheffer\}@u-bs.de}\and
Department of Computing Science, University of Oldenburg, Germany\\
\email{sven.von.hoeveling@uol.de}
}
\maketitle              
\begin{abstract}
We consider the online problem of packing circles into a square container.
A sequence of circles has to be packed one at a time, without knowledge
of the following incoming circles and without moving previously packed circles.
We present an algorithm that packs any online sequence of circles with a
combined area not larger than $0.350389$ of the square's area, improving the
previous best value of $\pi/10 \approx 0.31416$; even in an offline
setting, there is an upper bound of $\pi/(3+2\sqrt{2}) \approx 0.5390$. If only circles
with radii of at least $\radiusTwo$  are considered, our algorithm achieves the
higher value $0.375898$. 

As a byproduct, we give an online algorithm for packing circles into a
$1\times b$ rectangle with $b \ge 1$. This algorithm is worst case-optimal
for~$b \geq 2.36$. 

\end{abstract}
\section{Introduction}\label{sec:introduction}

Packing a set of circles into a given container is a natural geometric
optimization problem that has attracted considerable attention, 
both in theory and practice. 
Some of the many real-world applications are loading a shipping container
with pipes of varying diameter~\cite{George95}, packing paper products like
paper rolls into one or several containers~\cite{Fraser1994}, machine construction
of electric wires~\cite{Sugihara2004}, designing control
panels~\cite{Castillo2008}, placing radio towers with a maximal coverage
while minimizing interference~\cite{Szabo07}, industrial cutting~\cite{Szabo07},
and the study of macro-molecules or crystals~\cite{Peikert1994}. See the survey
paper of Castillo, Kampas, and Pint\'{e}r~\cite{Castillo2008} for an overview of other 
industrial problems. 
In many of these scenarios, the circles have to be packed
\emph{online}, i.e., one at a time, without the knowledge of further objects, 
e.g., when punching out a sequence of shapes from the raw material.

\begin{figure}[!htb]
	\centering
	\includegraphics[scale=0.52]{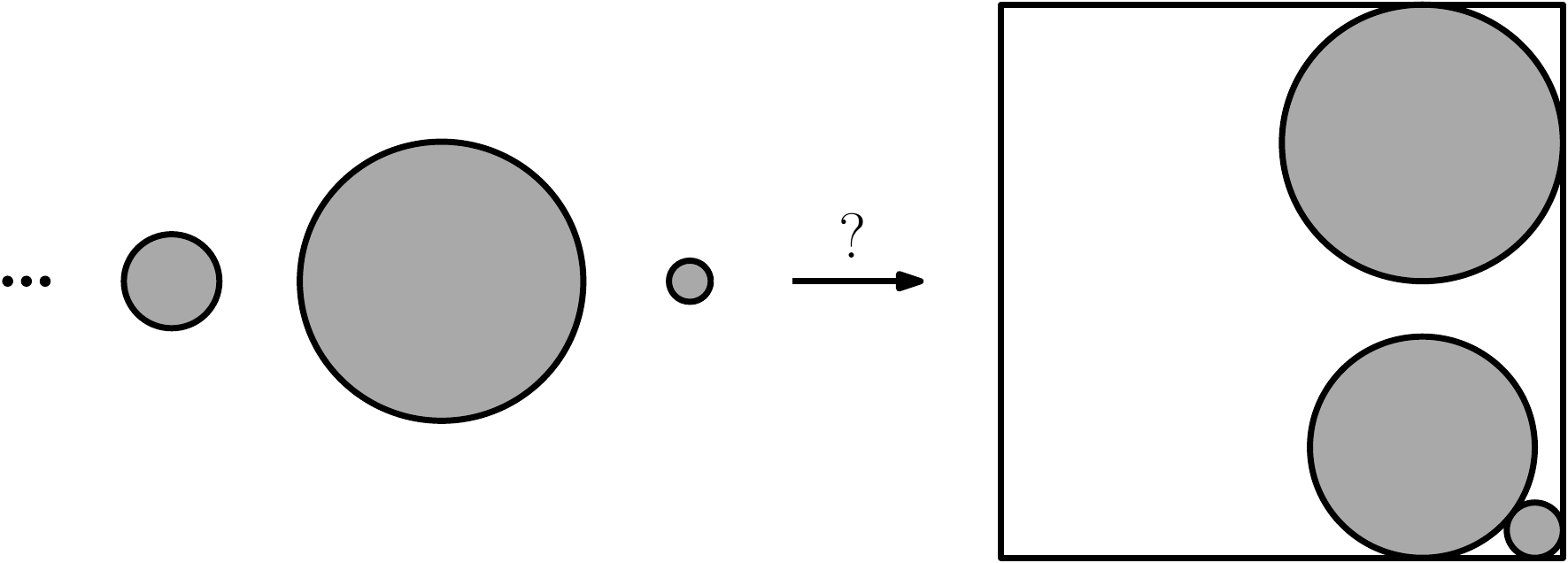}
	\caption[The problem]{Circles are arriving one at a time and have to be packed into the unit square. At this stage, the packed area is about $0.33$. What is the largest
$A \geq 0$ for which {\em any} sequence of total area $A$ can be packed?}
	\label{fig:problem}
\end{figure}


Even in an offline setting, deciding whether a given set of circles fits into
a square container is known to be NP-hard~\cite{Fekete10proceeding}, which is
also known for packing squares into a square~\cite{Leung90}. 
Furthermore, dealing with circles requires dealing with possibly
complicated irrational numbers, incurring very serious additional geometric
difficulties. This is underlined by the slow 
development of provably optimal packings of $n$ identical circles
into the smallest possible square.  
In 1965,
Schaer~\cite{schaer1965densest} gave the optimal solution for $n = 7$ and $n=8$
and Schaer and Meir~\cite{SM1965geometric} gave the optimal solution for $n =
9$. \new{A quarter of a century} later, W\"{u}rtz et al.~\cite{DPW1990optimal} provided
optimal solutions for $10$,$11$,$12$, and $13$ equally sized circles. In 1998,
Nurmela and Osterg{\aa}rd~\cite{NO1998more} provided optimal solutions for $n
\leq 27$ circles by making use of computer-aided optimality proofs. Mark{\'o}t
and Csendes~\cite{MC2005new} gave optimal solutions for $n = 28,29,30$ also by
using computer-assisted proofs within tight tolerance values. Finally, in 2002
optimal solutions were provided for $n \leq 35$ by Locatelli and
Raber~\cite{LR2002packing}; at this point, this is still the largest $n$ for
which optimal packings are known.
The extraordinary challenges of finding densest circle packings are also underlined
by a long-standing open conjecture by Erd\H{o}s and Oler from 1961~\cite{oler}
regarding optimal packings of $n$ unit circles into an equilateral triangle,
which has only been proven up to $n=15$. 

These difficulties make it desirable to develop relatively simple criteria
for the packability of circles. A natural bound arises from considering 
the {\em packing density}, i.e., the total area of objects compared to the size of the container;
the {\em critical packing density} $\delta$ is the largest value for 
which any set of objects of total area at most $\delta$ can be packed into a unit square;
see Fig.~\ref{fig:problem}.

\begin{figure}[!htb]
	\centering
	\subfigure{\includegraphics[height=0.325\textwidth]{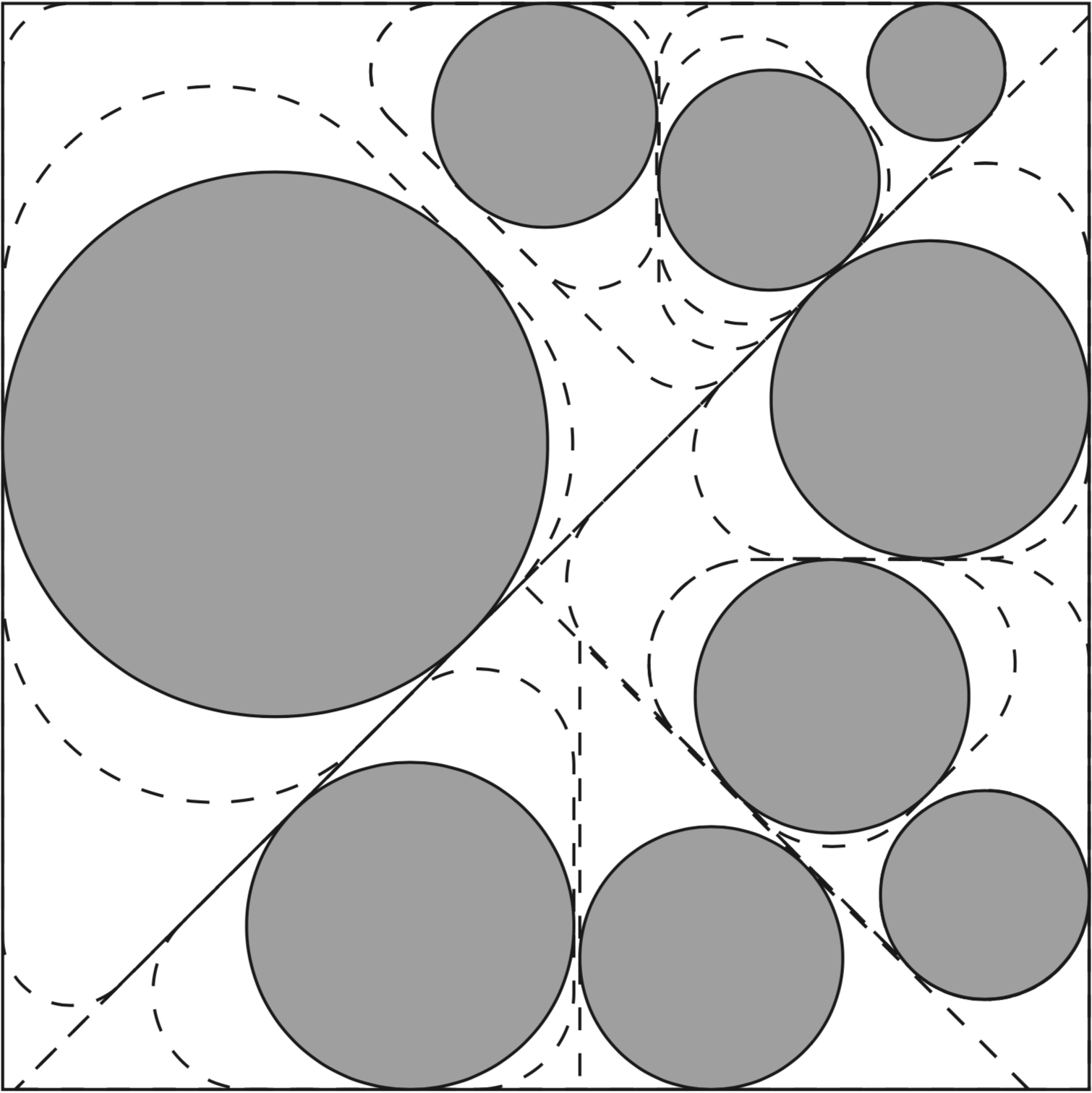}}
	\subfigure{\includegraphics[height=0.325\textwidth]{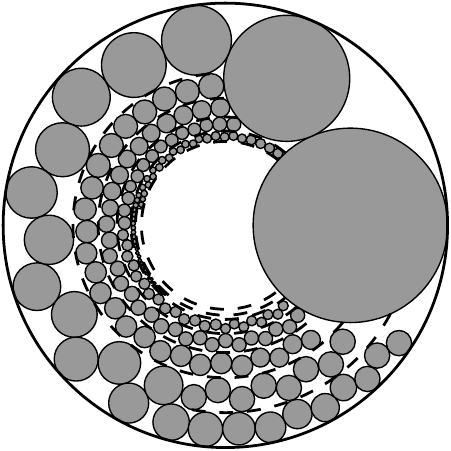}}
	\subfigure{\includegraphics[height=0.325\textwidth]{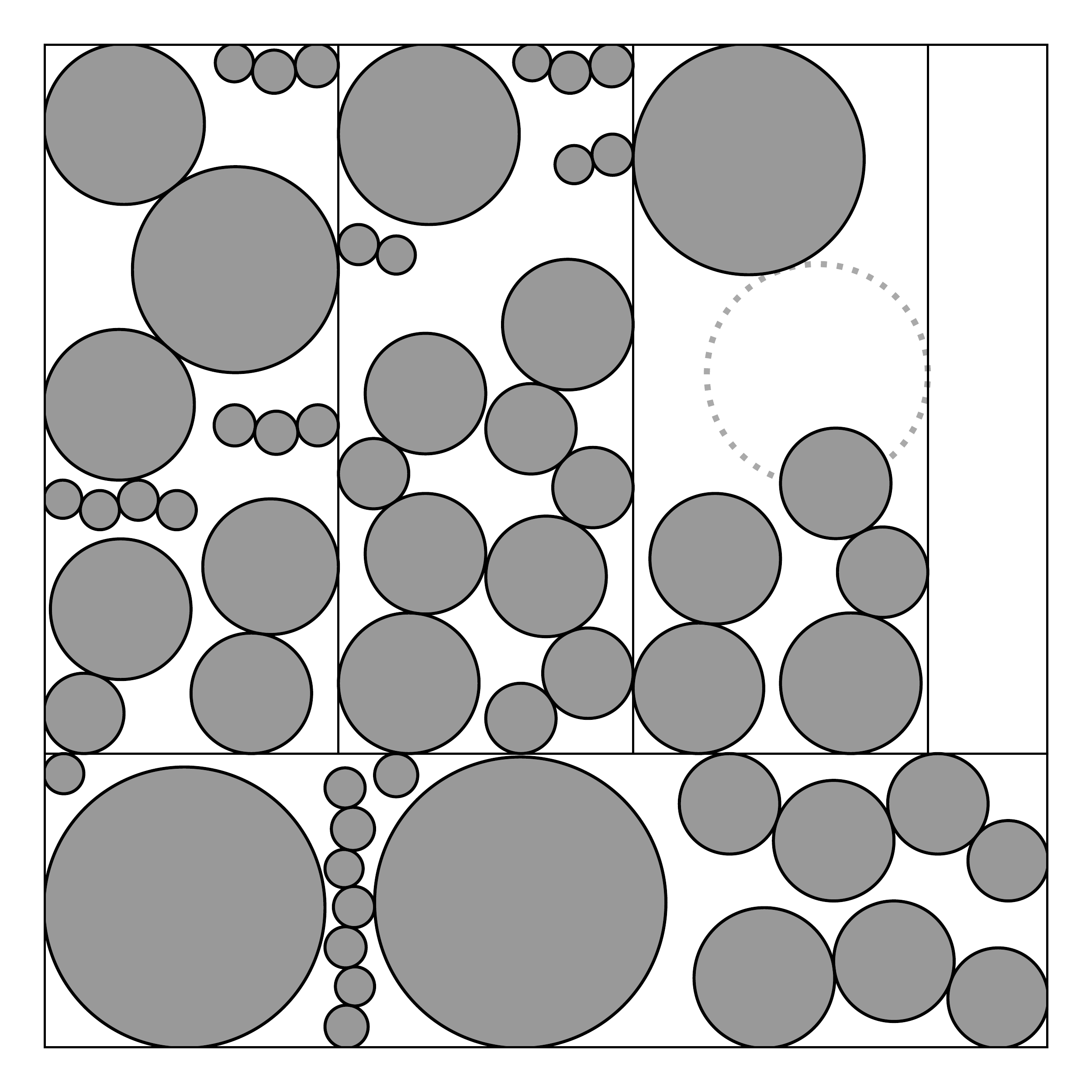}}
	\caption{Examples of algorithmic circle packings. 
(Left) The worst-case optimal offline algorithm of Fekete et al.~\cite{Fekete18} for packing
circles into the unit square. 
(Center) The worst-case optimal offline algorithm of Fekete et al.~\cite{fekete:circlesintocircles} for packing circles into the unit circle. 
(Right) Our online algorithm for packing circles into the unit square.} 
\label{fig:example}
\end{figure}

In an offline setting, two equally sized circles that fit exactly into
the unit square show that $\delta\leq \delta^*=\pi/(3+2\sqrt{2})\approx 0.5390$.
This is indeed tight: Fekete, Morr and Scheffer~\cite{Fekete18} gave a worst-case
optimal algorithm that packs any instance with combined area at most $\delta^*$;
see Fig.~\ref{fig:example}~(left).  More recently, Fekete, Keldenich and 
Scheffer~\cite{fekete:circlesintocircles} 
established $0.5$ as the critical packing density of circles in a circular container.

The difficulties of offline circle packing are compounded in an online setting. 
This is highlighted by the situation
for packing squares into a square, which does not encounter the mentioned
issues with irrational coordinates. It was shown in 1967 by 
Moon and Moser~\cite{Moser67} that the critical offline density is $0.5$:
Refining an approach by Fekete and Hoffmann~\cite{Fekete11}, Brubach~\cite{Brubach14}
established the currently best lower bound for online packing density of $0.4$.
This yields the previous best bound for the online packing density
of circles into a square: Inscribing circles into bounding boxes
yields a value of $\pi/10 \approx 0.3142$. 

\subsection{Our Results}

In this paper, we establish new lower bounds for
the online packing density of circles into \new{a square} and \new{into a rectangle}. \new{Note that in the online setting, a packing algorithm has to stop  as soon as it cannot pack a circle}.
We provide three online circle packing results for which we provide constructive proofs, i.e., corresponding algorithms guaranteeing the claimed packing densities. 

\begin{restatable}{theorem}{threctangleall}\label{th:rectangle_all}
	Let $b \geq 1$. Any online sequence of circles with a total area no larger than $\customMin\Big(0.528607 \cdot b - 0.457876,\frac{\pi}{4} \Big)$ can be packed into the $1 \times b$-rectangle $R$. This is worst-case optimal for $b\ge 2.36$.
\end{restatable}

We use the approach of Theorem~\ref{th:rectangle_all} as a subroutine and obtain the following:

\begin{restatable}{theorem}{thmmainsquare}\label{thm:mainsquare}
	Any online sequence of circles with a total area no larger than $0.350389$ can be packed into the unit square.
\end{restatable}

If the incoming circles' radii are lower bounded by $0.026623$, the density guaranteed by the algorithm of Theorem~\ref{thm:mainsquare} improves to $0.375898$.


\begin{restatable}{theorem}{thmmainsquarenotiny}\label{thm:mainsquare_notiny}
	Any online sequence of circles with radii not smaller than $0.026623$ and with a total area no larger than $0.375898$ can be packed into the unit square.
\end{restatable}




We describe the algorithm of Theorem~\ref{th:rectangle_all} in Section~\ref{sec:algorithm} and the algorithm of the Theorems~\ref{thm:mainsquare} and~\ref{thm:mainsquare_notiny} in Section~\ref{sec:algorithm_unit_square}.


\section{Packing into a Rectangle}\label{sec:algorithm}
In this section, we describe the algorithm, \emph{Double-Sided Structured Lane Packing (DSLP)}, of Theorem~\ref{th:rectangle_all}. In particular, DSLP uses a packing strategy called \emph{Structured Lane Packing (SLP)} and an extended version of SLP as subroutines.

\subsection{Preliminaries for the Algorithms}\label{sec:preliminiaries}

A \emph{lane} $L \subset \mathbb{R}^2$ is an $x$- and $y$-axis-aligned rectangle. The \emph{length} $\ell(L)$ and the \emph{width} $w(L)$ of $L$ are the dimensions of $L$ such that $w(L) \leq \ell(L)$. $L$ is \emph{horizontal} if the length of $L$ is given via the extension of $L$ w.r.t. the $x$-axis. Otherwise, $L$ is \emph{vertical}. The \emph{distance} between two circles packed into $L$ is the distance between the \new{orthogonal projections of the circles' midpoints onto the longer side of $L$.} A lane is either $\emph{open}$ or \emph{closed}. Initially, each lane is open. 

\emph{Packing} a circle $C$ into a lane $L$ means placing $C$ inside $L$ such
that $C$ does not intersect another circle that is already packed into~$L$
or into another lane. A \emph{(packing) strategy} for a lane $L$ is a set of
rules that describe how a circle has to be packed into $L$. The
\emph{(packing) orientation} of a strategy for a horizontal lane is either
\emph{rightwards} or \emph{leftwards} and the \emph{(packing) orientation} of a
strategy for a vertical lane is either \emph{downwards} or \emph{updwards}.


Let $w$ be the width of $L$. Depending on the radius $r$ of the current circle $C$, we say: $C$ is \emph{medium} (Class 1)  if $w > r \geq \frac{w}{4}$, $C$ is small if $\frac{w}{4} > r \geq 0.0841305w$ (Class 2), $C$ is \emph{tiny} (Class 3 or 4) if $0.0841305w > r \geq 0.023832125w$, and $C$ is \emph{very tiny} if $0.023832125w > r$ (Classes 5,6, \dots). For a more refined classification of $r$, we refer to Section~\ref{sec:fillinggaps}. \new{The general idea is to reach a certain density within a lane by packing only relatively equally sized circles into a lane with SLP.}

For the \new{rest} of Section~\ref{sec:algorithm}, for $0 < w \leq b$, let $L$ be a horizontal $w \times b$ lane. 

\subsection{Structured Lane Packing (SLP) \new{--} The Standard Version}\label{sec:lanepacking}

\begin{figure}[t]
	\centering
	\includegraphics[scale=0.7]{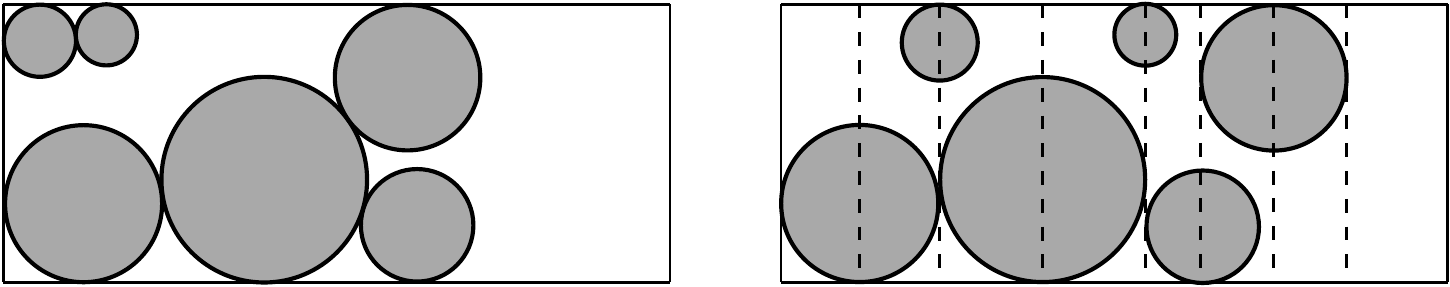}
	\caption[Tight Lane Packing compared with Structured Lane Packing]{Comparison of Tight Lane Packing (left) with Structured Lane Packing (right) for the same input. The former has a smaller packing length.}
	\label{fig:comptightstructuredlanepacking}
\end{figure}

Rightwards \emph{Structured Lane Packing (SLP)} packs circles into $L$ alternating touching the bottom and the top side of $L$ from left to right, see Fig.~\ref{fig:comptightstructuredlanepacking} (right).

In particular, we pack a circle $C$ into $L$ as far as possible to the left while guaranteeing: (1) $C$ does not overlap a vertical lane packed into $L$, see Section~\ref{sec:fillinggaps} for details\footnote{Requiring that $C$ does not overlap a vertical lane placed inside $L$ is only important for the extension of SLP (see Section~\ref{sec:fillinggaps}), because the standard version of SLP does not place vertical lanes inside $L$.}. (2) The distance between $C$ and the circle~$C'$ packed last into $L$ is at least $\min \{ r,r' \}$ where $r,r'$ are the radii of $C,C'$, see Fig.~\ref{fig:comptightstructuredlanepacking} (right). 

Leftwards \emph{Structured Lane Packing} packs circles by alternatingly touching
the bottom and the top side of $L$ from right to left.
Correspondingly, upwards and downwards \emph{Structured Lane Packing} packs
circles alternatingly touching the left and the right side of $L$ from bottom to top
and from top to bottom.


\subsection{Extension of SLP \new{--} Filling Gaps by (Very) Tiny Circles}\label{sec:fillinggaps}

\begin{figure}[!htb]
	\centering
	\includegraphics[height=7cm]{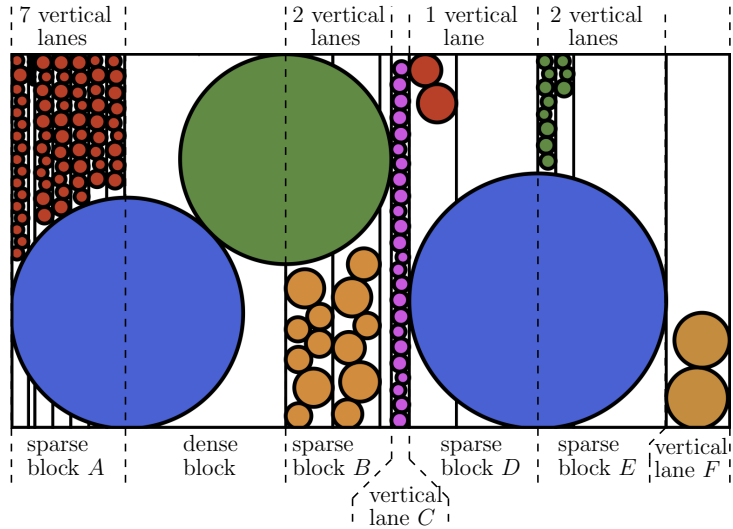}
	
	\caption[Circle centre partition]{A packing produced by extended SLP: 128  (3 medium, 17 tiny, and 108 very tiny) circles packed into 15 (1 medium, 4 tiny and 10 very tiny) lanes. A possible input order of the circles is \textcolor{blue}{1 medium circle}, \textcolor{red}{23 very tiny circles} (filling the sparse block $A$), \textcolor{seagreen}{1 medium circle}, \textcolor{orange}{13 tiny circles} (filling the sparse block $B$), \textcolor{violet}{24 very tiny circles} (filling the vertical lane $C$), \textcolor{blue}{1 medium circle}, \textcolor{red}{2 tiny circles} (filling the sparse block $D$), \textcolor{seagreen}{11 very tiny circles} (filling the sparse block $E$), and \textcolor{orange}{2 tiny circles} (filling the vertical lane $F$).}
	\label{fig:partitions}
\end{figure}

Now consider packing medium circles with SLP. We extend SLP for placing tiny and very tiny circles within the packing strategy, see Fig.~\ref{fig:partitions}. Note that small circles are not considered for the moment, such that (very) tiny circles are relatevily small compared to the medium ones. In particular, if the current circle~$C$ is medium, we apply the standard version of SLP, as described in Section~\ref{sec:lanepacking}. If~$C$ is (very) tiny, we pack $C$ into a vertical lane inside $L$, as described next.

We pack (very) tiny circles into vertical lanes inside $L$, see Fig.~\ref{fig:partitions}. The vertical lanes are placed inside \emph{blocks} that are the rectangles induced by vertical lines touching medium circles already packed into $L$, see Fig.~\ref{fig:comptightstructuredlanepacking} (right).

Blocks that include two halves of medium circles are called \emph{dense} blocks,
while blocks that include one half of a medium circle are called \emph{sparse}
blocks. The area of $L$ that is neither covered by a dense block or a sparse
block is called a \emph{free} block. \emph{Packing} a vertical lane $L'$ into a
sparse block~$B$ means placing $L'$ inside~$B$ as far as possible to the left,
such that~$L'$ does not overlap another vertical lane already packed
into~$B$. Packing a vertical lane~$L'$ into $L$ means placing~$L'$ inside~$L$
as far as possible to the left, such that~$L'$ does neither overlap another
vertical lane packed into~$L$, a dense block of~$L$, or a sparse block of $L$.

We extend our classification of circles by defining classes $i$ of lane widths $w_i$ and \new{\emph{relative lower bounds} $q_i$} for the circles' radii as described in Table~\ref{table:lane_sizes}. This means a circle with radius $r$ belongs \new{to class 1 if $0.5w \geq r > q_1 w_1$ and to class $i$ if $q_{i-1} w_{i-1} \geq r > q_{i} w_{i}$}, for $i \geq 2$.
Only circles of class $i$ are allowed to be packed into lanes of class $i$.

\begin{table}[!htb]
	\centering
	\begin{tabularx}{\textwidth}{ p{2.4cm} p{4.2cm}  X }\toprule
		\textbf{Class $i$ } & \textbf{\new{(Relative)} lower bound $q_i$}  & \textbf{Lane width $w_i$}  \\ \midrule 
		1 (Medium) & $q_1\coloneqq \qM \coloneqq 0.25$ & $w_1 \coloneqq w$ \\ \hdashline
		2 (Small) & $q_2 \coloneqq \qS \coloneqq 0.168261$  & $w_2\coloneqq2 \qM  w= 0.5 \cdot w$   \\ \hdashline
		3 (Tiny) & $q_3 \coloneqq 0.371446$ &  $w_3\coloneqq4  \qM \qS w= 0.168261 \cdot w$  \\ \hdashline
		4 (Tiny) & $q_4 \coloneqq 0.190657$ & $w_4\coloneqq8  \qM \qS q_3 w\approx 0.125 \cdot w$  \\ \hdashline
		5 (Very tiny) & $q_5 \coloneqq 0.175592$ &  $w_5\coloneqq16  \qM \qS q_3 q_4 w\approx 0.047664 \cdot w$  \\ \hdashline
		6 (Very tiny) & $q_6 \coloneqq 0.170699$ &  $w_6\coloneqq32 \qM \qS q_3 q_4 q_5 w\approx 0.016739 \cdot w$  \\ \hdashline
		7 (Very tiny) & $q_7 \coloneqq 0.169078$ &  $w_7\approx 0.005715 \cdot w$  \\ \hdashline
		8 (Very tiny) & $q_8 \coloneqq 0.168354$ &  $w_8\approx 0.001932 \cdot w$  \\ \hdashline
		9 (Very tiny) & $q_9 \coloneqq 0.168293$ &  $w_9\approx 0.000651 \cdot w$  \\ \hdashline
		10 (Very tiny) & $q_{10} \coloneqq 0.168272$ &  $w_{10}\approx 0.000219 \cdot w$  \\ \hdashline
		11 (Very tiny) & $q_{11} \coloneqq 0.168265$ &  $w_{11}\approx 0.000074 \cdot w$  \\ \hdashline
		12 (Very tiny) & $q_{12} \coloneqq 0.168263$ &  $w_{12}\approx 0.000025 \cdot w$  \\ \hdashline
		13 (Very tiny) & $q_{13} \coloneqq 0.168262$ &  $w_{13}\approx 0.000008 \cdot w$  \\ \hdashline
		\dots & $\dots$ &  $\dots$ \\ \hdashline
		k (Very tiny) & $q_k \coloneqq 0.168262$ & $w_k\coloneqq 2^{k-1}  \qM \qS q_3 q_4 \cdot \ldots \cdot q_{k-1} w$  \\ \bottomrule
	\end{tabularx}
	\caption[Lower circle size limits and lane widths]{Circles are classified into the listed classes. Note that the lower bounds to the circles' radii is relative to the lane width, e.g., the absolute lower bound for circles inside a small lane is $q_S w_2 = 0.168261w_2$.} \label{table:lane_sizes}
\end{table}

A sparse block is either \emph{free}, \emph{reserved} for class $3$, \emph{reserved} for class $4$, \emph{reserved} for all classes $i \geq 5$, or \emph{closed}. Initially, each sparse block is free. 

We use SLP in order to pack a circle $C$ of class $i \geq 3$, into a vertical lane $L_i \subset L$ of class $i$ and width $w_i$ by applying the Steps~1-5 in increasing order as described below. When one of the five steps achieves that $C$ is packed into a vertical lane~$L_i$, the approach stops and returns successful.

\begin{itemize}
	\item \textbf{Step (1):} If there is no open vertical lane $L_i \subset L$ of class $i$ go to Step~2. 	Assume there is an open vertical lane $L_i$ of class $i$. If $C$ can be packed into~$L_i$, we pack $C$ into $L_i$. Else, we declare $L_i$ to be closed.
	
	\item \textbf{Step (2):} We close all sparse blocks $B$ that are free or reserved for class~$i$ in which a vertical lane of class $i$ cannot be packed into $B$. 
	
	\item \textbf{Step (3):} If there is an open sparse block $B$ that is free or reserved for class~$i$ and a vertical lane of class $i$ can be packed into $B$:
	\begin{itemize}
		\item \textbf{(3.1):} We pack a vertical lane $L_i \subset L$ of class~$i$ into $B$. If the circle half that is included in $B$ touches the bottom of $L$, we apply downwards SLP to~$L_i$. Otherwise, we apply upwards SLP to~$L_i$.
		\item \textbf{(3.2):} If~$B$ is free and $i \in \{ 3,4 \}$, we reserve $B$ for class $i$. If~$B$ is free and~$i \geq 5$, we reserve $B$ for all classes $i \geq 5$.
		\item \textbf{(3.3):} We pack $C$ into~$L_i$.
	\end{itemize}
	\item \textbf{Step (4):} If a vertical lane of class $i$ can be packed into $L$: 
		\begin{itemize}
			\item \textbf{(4.1):} We pack a vertical lane $L_i$ of class $i$ into $L$ and apply upwards~SLP to $L_i$.
			\item \textbf{(4.2):} We pack $C$ into $L_i$
		\end{itemize}
	\item \textbf{Step (5):} We declare $L$ to be closed and return failed.
\end{itemize}

\subsection{Double-Sided Structured Lane Packing (DSLP)}

\begin{figure}[h!]
	\centering
	\includegraphics[scale=0.35]{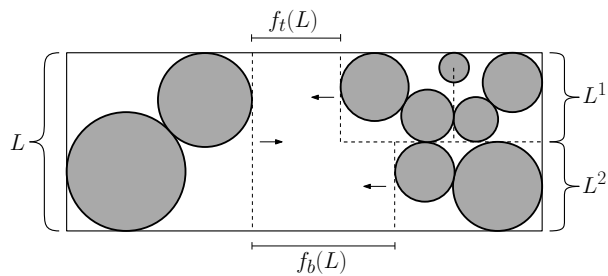}
	\caption[A packing of a medium lane (from left to right) and of the two contained small lanes (from right to left)]{A packing produced by DSLP: The medium lane is packed from left to right by medium circles. The two contained small lanes are packed simultaneously in parallel from right to left by small circles.}\label{fig:two_sided_packing}
\end{figure}

We use SLP as a subroutine in order to define our packing strategy \emph{Double-Sided Structured Lane Packing (DSLP)} of Theorem~\ref{th:rectangle_all}. In particular, additionally to $L$, we consider two small lanes $L^1,L^2$ that partition~$L$, see Fig.~\ref{fig:two_sided_packing}.

Rightwards \emph{Double-Sided Structured Lane Packing (DSLP)} applies the extended version of rightwards SLP to $L$ and leftwards SLP to $L^1,L^2$. If the current circle~$C$ is medium or (very) tiny, we pack $C$ into $L$. If $C$ is small, we pack $C$ into that lane of $L^1,L^2$, resulting in a smaller packing length. 

Leftwards DSLP is defined analogously, such that the extended version of leftwards SLP is applied to $L$ and rightwards SLP to $L^1,L^2$. Correspondingly, upwards and downwards DSLP are defined for vertical lanes.

\section{Packing into the Unit Square}\label{sec:algorithm_unit_square}
We extend our circle classification by the class $0$ of \emph{large} circles and define a relative lower bound $q_0:= \frac{w}{2}$ and the lane width of corresponding \emph{large} lanes as $w_0 := 1-w$.

\new{We set $w$ to $0.288480$ and $0.277927$ for Theorem~\ref{thm:mainsquare} respectively Theorem~\ref{thm:mainsquare_notiny}.}
In order to pack large circles, we use another packing strategy called \emph{Tight Lane Packing (TLP)} defined as SLP, but \new{without restrictions} (1) and (2), see Fig.~\ref{fig:comptightstructuredlanepacking}.

\begin{figure}[h!]
	\centering
	\subfigure{\includegraphics[width=0.4\textwidth]{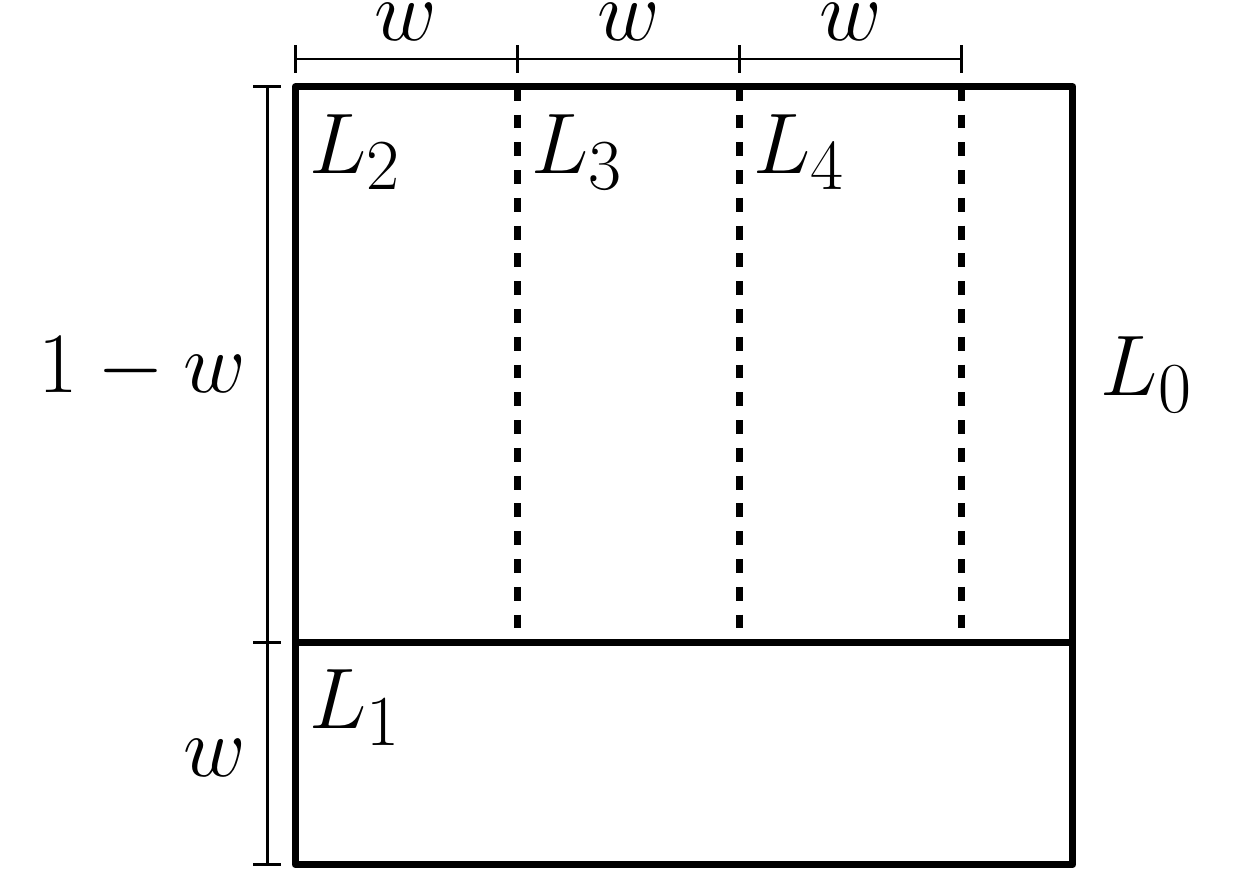}}
	\hspace{1cm}
	\subfigure{\raisebox{-1.6mm}{\includegraphics[width=0.276\textwidth]{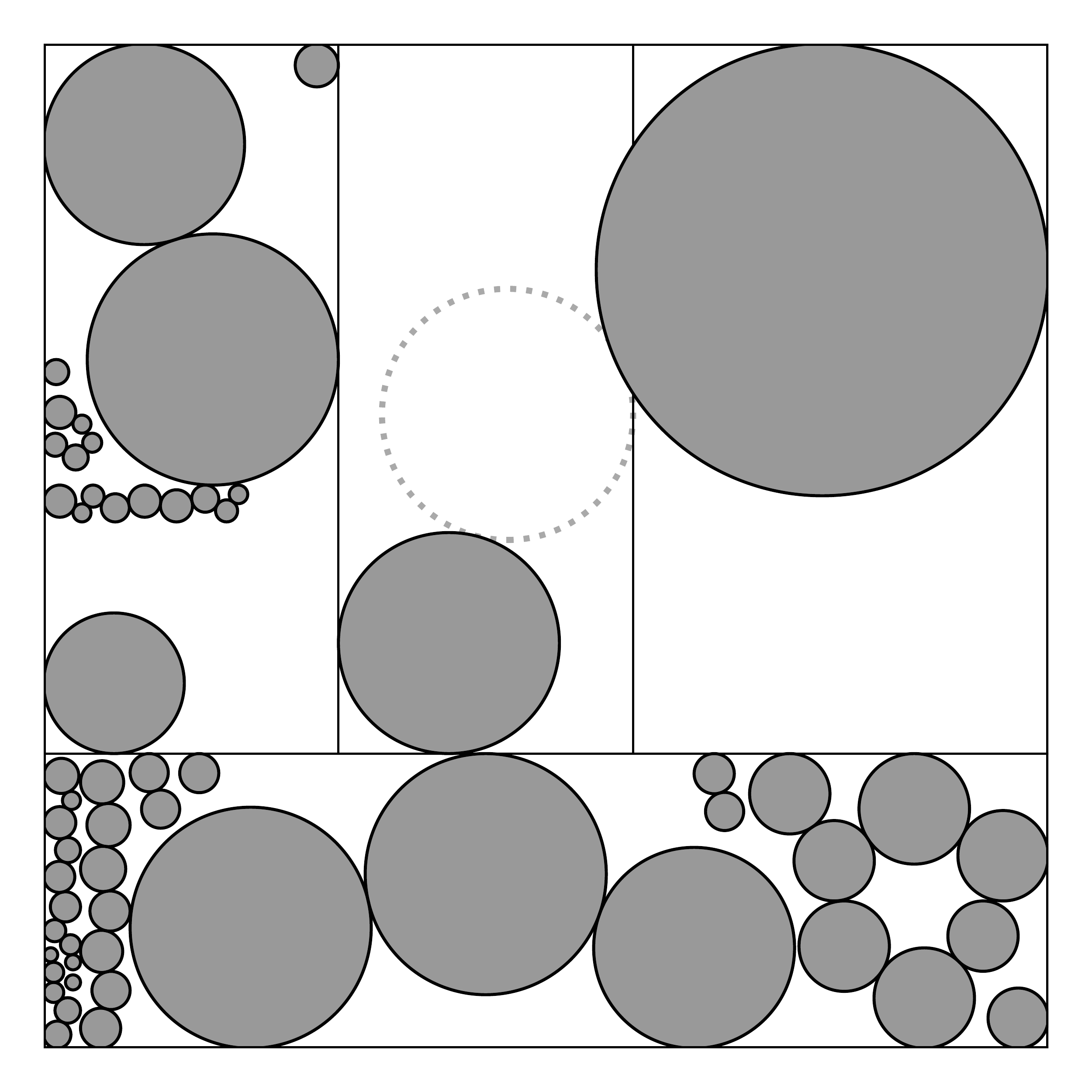}}}
	\caption[Partition of the unit square]{Left: The unit square is divided into four lanes $L_1$, $L_2$, $L_3$, and $L_4$, into which medium, small, tiny, and very tiny circles are packed. Large circles are packed into a lane $L_0$ that overlaps $L_1$, $L_2$, and $L_3$. Right: An example packing. A medium circle (dotted) does not fit.}
	\label{fig:partition}
\end{figure}

We cover the unit square by the union of one large lane $L_0$ and four medium lanes $L_1,\dots,L_k$ for $k=4$, see Fig.~\ref{fig:partition}. We apply TLP to $L_0$ and DSLP to $L_1,\dots,L_4$. The applied orientations for~$L_0,L_1,L_2$ are leftwards, rightwards, and downwards. For $i = 3,4$, the orientation for $L_i$ is chosen such that the first circle packed into $L_i$ is placed adjacent to the bottom side of $L_i$.

If the current circle to be packed is large, we pack $C$ into the large lane $L_0$ and stop if $C$ does not fit in $L_0$. Otherwise, in increasing order we try to pack~$C$ into $L_1,\dots,L_4$.

\section{Analysis of the Algorithms}\label{sec:analysis}
In this section we sketch the analysis of our approaches and refer to the appendix for full details. First, we analyze the packing density guaranteed by DSLP. Based on that, we prove our main results Theorems~\ref{th:rectangle_all},~\ref{thm:mainsquare}, and~\ref{thm:mainsquare_notiny}.

\subsection{Analysis of SLP}

In this section, we provide a framework for analyzing the packing density guaranteed by DSLP for a horizontal lane $L$ of width $w$. \new{It is important to note that this framework and its analysis in this subsection deals with the packing of only one class into a lane.} 

\new{We introduce some definitions.} The \emph{packing length} $\pl(L)$ is the maximal difference of $x$-coordinates of \new{points from} circles packed into $L$. The \emph{circle-free length} $\fl(L)$ of $L$ is defined as $\len(L) - \pl(L)$.
\new{We denote the \textit{total area} of a region $R \subset \mathbb{R}^2$ by $\area(R)$ and the area of an $a \times b$-rectangle by $\R(a, b)$. Furthermore, we denote the area of a semicircle for a given radius $r$ by $\mathcal{\HC}(r)\coloneqq \frac{\pi}{2} r^2$. The total area of the circles packed into~$R$ is called \textit{occupied area} denoted by $\occ(R)$.
Finally, the density $\den(R)$ is defined by $\occ(R)/\area(R)$.}

In order to apply our analysis for different classes of lanes, i.e., different lower bounds, we consider a general \new{(relative) lower bound} $q$ for the radii of circles allowed to be packed into $L$ with $0 < q \le 1/2$. The following lemma deduces a lower bound for the density of dense blocks depending on $q$.

\begin{restatable}{lemma}{thpackingtwo}\label{th:packing_two}
	Consider a dense block $D$ containing two semicircles of radii $r_1$ and~$r_2$ such that $0<\rlow \w\le r_1,r_2\le1/2\w$. \new{Then $\den(D)$ is lower-bounded by} 
	\begin{equation*}
	\dlow: \left(0, \frac{1}{2}\right] \rightarrow \mathbb{R}  \text{ with } 
	\rlow \mapsto
	\begin{cases}
	\pi \rlow & 0 < \rlow < \frac{1}{3\sqrt{3}} \\
	\frac{\pi}{3\sqrt{3}} \approx 0.6046 & \frac{1}{3\sqrt{3}} \le \rlow \le \frac{1}{3}\\
	\frac{\pi \rlow^2}{\sqrt{4\rlow-1}} & \frac{1}{3} < \rlow \le \frac{1}{2}.
	\end{cases}
	\end{equation*}
\end{restatable}

\begin{figure}[!h]
	\centering
	\subfigure{\includegraphics[scale=0.37]{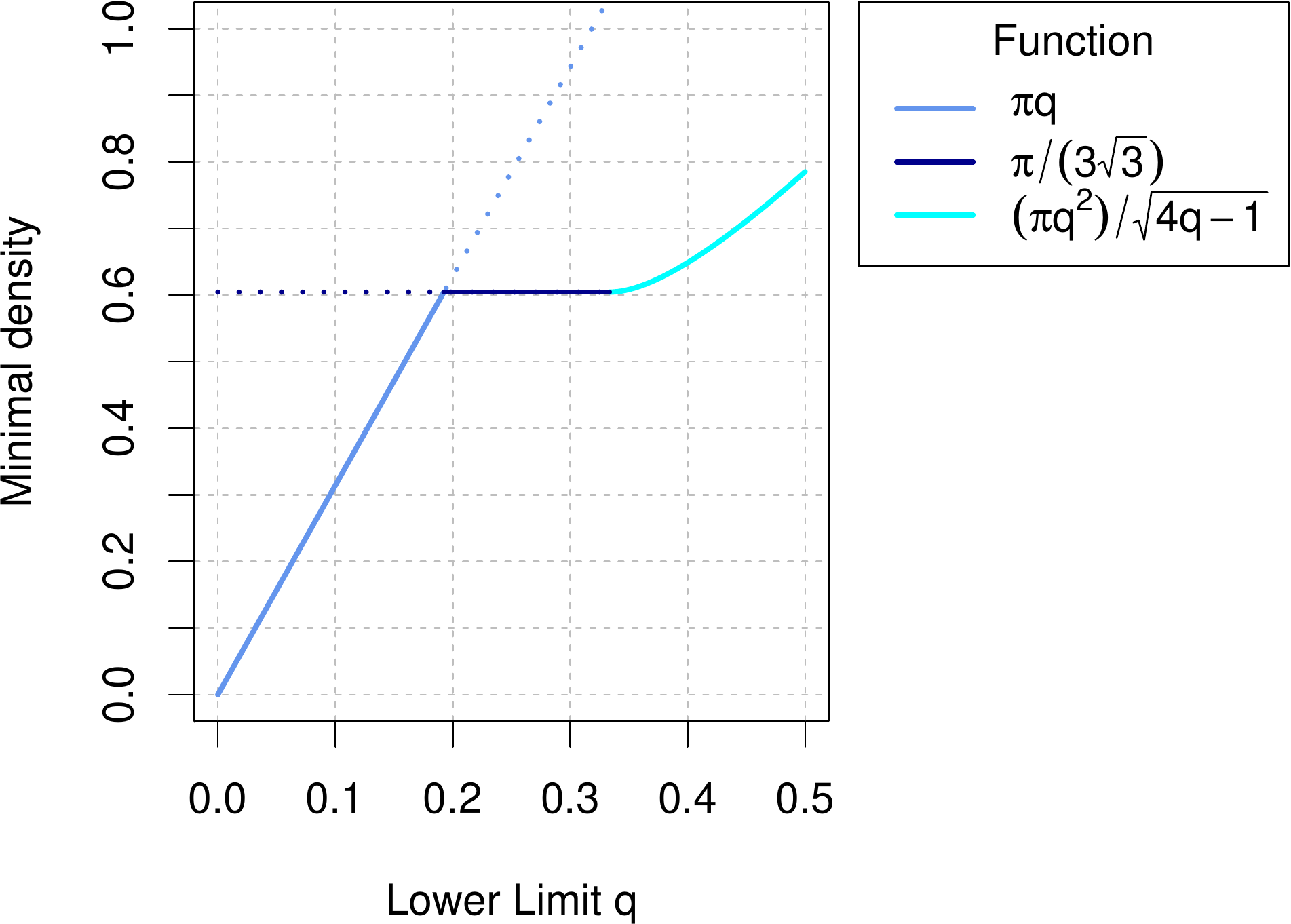}}
	\hspace{4ex}
	\subfigure{\includegraphics[scale=0.36]{figures/7_rechts}}
	\caption[Density bound function $\dlow$ for dense blocks]{(Left): A plot
of $\dlow(q)$ for its complete range. It provides the minimal density of a
dense block whose two semicircles have a radius of at least $q \cdot w$. A
lower bound of $q=1/2$ leads to a minimal density of $\pi/4$ which is the ratio
of a circle to its minimal bounding square. (Right):  (1): $\delta(0.15) \approx
0.47123$, (2): $\delta(\frac{1}{3 \sqrt{3}}) \approx 0.6046$, and (3):
$\delta(0.4) \approx 0.6489$.}\label{fig:density_function_main_part} \end{figure} 	


We continue with the analysis of sparse blocks. Sparse blocks have a minimum length $qw$. 
\hyperref[th:sparse_block]{Lemma~\ref*{th:sparse_block}} states a lower bound for the occupied area of sparse blocks. This lower bound consists of a constant summand and a summand that is linear with respect to the actual length. 

\begin{restatable}{lemma}{thsparseblock}\label{th:sparse_block}
Given a density bound $\dmin \le \dlow(q)$ for dense blocks. Let $S$ be a sparse block and $z$ be the lower bound for $\len(S)$ with \mbox{$\len(S) \ge z \ge qw$}. Then $\occ(S) \ge \R(\len(S) - z, w) \cdot \dmin + \HC(z)$.
\end{restatable}

\begin{figure}[!h]
	\centering
 	\includegraphics[scale=0.34]{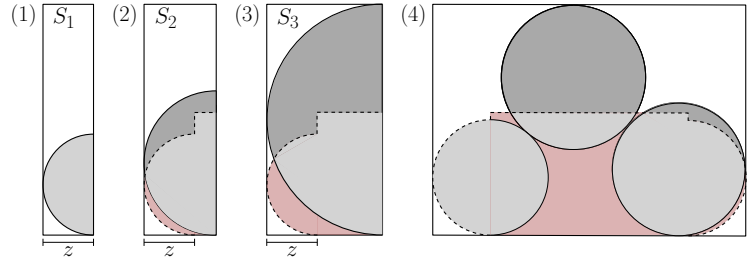}
	\caption[Illustration of the bound for sparse blocks]{(1)+(2)+(3)
Three sparse blocks $S_1, S_2,S_3$ in order of ascending length. The grey
coloured area (light and dark unified) represents the occupied area. The dashed
area shows the lower bound of
\hyperref[th:sparse_block]{Lemma~\ref*{th:sparse_block}}, which is composed of
the smallest possible semicircle plus a linear part. The dark grey parts
symbolize the area that exceeds the bound, whereas the red parts symbolize the
area missing to the bound. Block $S_1$ has the minimal length $p$ so that the
occupied area and the bound are equal. For blocks of larger length,
represented by $S_2$ and $S_3$, the dark grey area is larger than the red area.
(4) A packing produced by SLP and the lower bound of
\new{\hyperref[th:min_sp]{Lemma~\ref*{th:min_sp}}.}}
\label{fig:sparse_block2}
\end{figure}

The occupied area of a sparse block is at least a semicircle of a smallest possible circle plus the remaining length multiplied by the lane width and by the minimal density $\dmin$ of dense blocks. This composition is shown in \hyperref[fig:sparse_block]{Fig.~\ref*{fig:sparse_block2}}.

\new{Next, we combine the results of Lemma~\ref{th:packing_two} and Lemma~\ref{th:sparse_block}. We define the term $\M\big(p,w,z,\dmin\big)\coloneqq \R(p-2z,w) \cdot \dmin + 2 \cdot \HC(z)$ for some $p,w,z,\dmin > 0$ and state the following (see also \hyperref[fig:sparse_block]{Fig.~\ref*{fig:sparse_block2}} (4)).}


\new{
\begin{restatable}{lemma}{thminsp}\label{th:min_sp}
	Given a lane $L$ packed by \SP, a lower bound $q$, and a density bound $\dmin \le \dlow(q)$ for dense blocks. Let $w$ be the width of $L$. 
	The occupied area in $L$ is lower-bounded by $\M\big(\pl(L),w,qw,\dmin\big)$.
\end{restatable}
}

\subsection{Analysis of DSLP}\label{sec:analysisDSLP}

Let $L$ be a horizontal lane packed by DSLP. We define $\plt(L)$ ($\plb(L)$) as the sum of the packing lengths of the packing inside $L$ and the length of the packing inside the top (bottom) small lane inside $L$. Furthermore, we define $\flt(L) := \ell(L) - \plt(L)$ and $\flb(L) := \ell(L) - \plb(L)$, see Fig.~\ref{fig:two_sided_packing}. 

By construction, vertical dense blocks packed into $L$ have a density of at least $\delta(q_2)$. In fact, the definitions of circle sizes for all classes $i \geq 2$ ensure the common density bound $\dstar := \delta(q_2)$ for dense blocks.

We consider \new{mixed dense blocks, that were defined as sparse blocks of $L$ in which vertical lanes are packed, also as dense blocks and extend the lower bound $\den(D)\ge \dstar$ to all kinds of dense blocks by the following Lemma. }

\begin{restatable}{lemma}{thdenseblockbound}\label{th:dense_block_bound}
Let $D$ be a dense block of $L$. Assume all vertical lanes packed into $D$ to be closed. Then $\den(D)\ge \dstar$.
\end{restatable}

As some vertical lanes may not be closed, we upper bound the error $\OES(L)$ that we make by assuming that all vertical lanes $L_1,\dots,L_n \subset L$ are closed.

\begin{restatable}{lemma}{thoes}\label{th:oes}
$\OES(L_1 \cup \ldots \cup L_n) < 0.213297 \cdot w^2$.
\end{restatable}

We lower bound the occupied area inside $L$ by using the following term:
\begin{eqnarray*}
	\MDSP(\plt(L),\plb(L),w,z, \dmin):=&& \R\big(\plt(L)+\plb(L)- w - 4z,w_2\big) \cdot \dmin 
 \\
 	&&+ 2 \cdot  \HC\big(\frac{w}{4}\big) + 4 \HC(z),
\end{eqnarray*}
where $z$ denotes the minimal radius for the circles, i.e., $z := q_2 w_2$.

\begin{figure}[h!]
	\centering
	\includegraphics[scale=0.45]{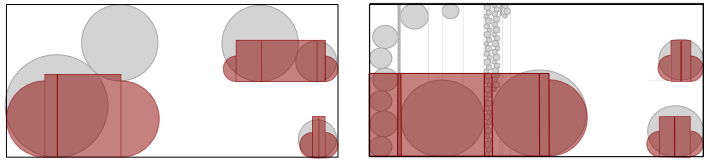}
	\caption{Two example packings and the compositions of our lower bounds (red) for the occupied area implied by Lemma~\ref{th:one_lane_advanced}. \new{Note that $\OES(L)$ is not visualized. }}
	\label{fig:vis_lower_bound}
\end{figure}

\new{Applying Lemma~\ref{th:dense_block_bound} and Lemma~\ref{th:sparse_block} separately to $L$, $L^1$, and $L^2$, analogous to the combination of Lemma~\ref{th:packing_two} and Lemma~\ref{th:sparse_block} in the last subsection, and estimating the error $\OES(L)$ with Lemma~\ref{th:oes}, yields lower bounds for the occupied areas of $L$, $L^1$, and $L^2$. Fig.~\ref{fig:vis_lower_bound} separately illustrates the lower bounds for the occupied areas of $L$, $L^1$, and $L^2$ for two example packings and Lemma~\ref{th:one_lane_advanced} states the result.}

\vspace{-5ex}




\begin{restatable}{lemma}{thonelaneadvanced}\label{th:one_lane_advanced}
$\MDSP\big(\pl_t(L),\pl_b(L),w,z,\dstar \big) - \OES(L) \geq \occ(L)$.
\end{restatable}

\subsection{Analysis of Packing Circles into a Rectangle}

Given a \new{$1\times b$} rectangle $R$, we apply DSLP for packing the input circles into $R$. 

\restatethm{\threctangleall*}{th:rectangle_all}

\begin{figure}[h!]
	\centering
	\includegraphics[scale=0.285]{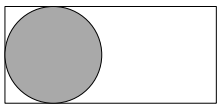}
	\caption{\new{A} worst case for packing circles into an $1 \times b$-rectangle \new{with $b\ge 2.36$} consists of one circle with \new{radius $\frac{1}{2} + \epsilon$}. \new{The shown circle with radius $\frac{1}{2}$ just fits.}}
	\label{fig:worst_case}
\end{figure}

The lower bound for the occupied area implied by Lemma~\ref{th:one_lane_advanced} is equal to $\big(b-\frac{3}{4}-q_2) \cdot \dstar + \frac{\pi}{16} + \frac{\pi}{2}(q_2)^2 - 0.213297$. This is lower bounded by $\frac{\pi}{4}$ for $b \geq 2.36$. \new{Hence, the online sequence consisting of one circle with a radius of $\frac{1}{2} + \epsilon$ and resulting total area of $\frac{\pi}{4} + \epsilon$ is a worst case online sequence for $b\ge 2.36$}, see Fig.~\ref{fig:worst_case}. This concludes the proof of Theorem~\ref{th:rectangle_all}.

\vspace{-4ex}

\subsection{Analysis of Packing Circles into the Unit Square}

\begin{figure}[!ht]
	\centering
	\includegraphics[scale=0.29]{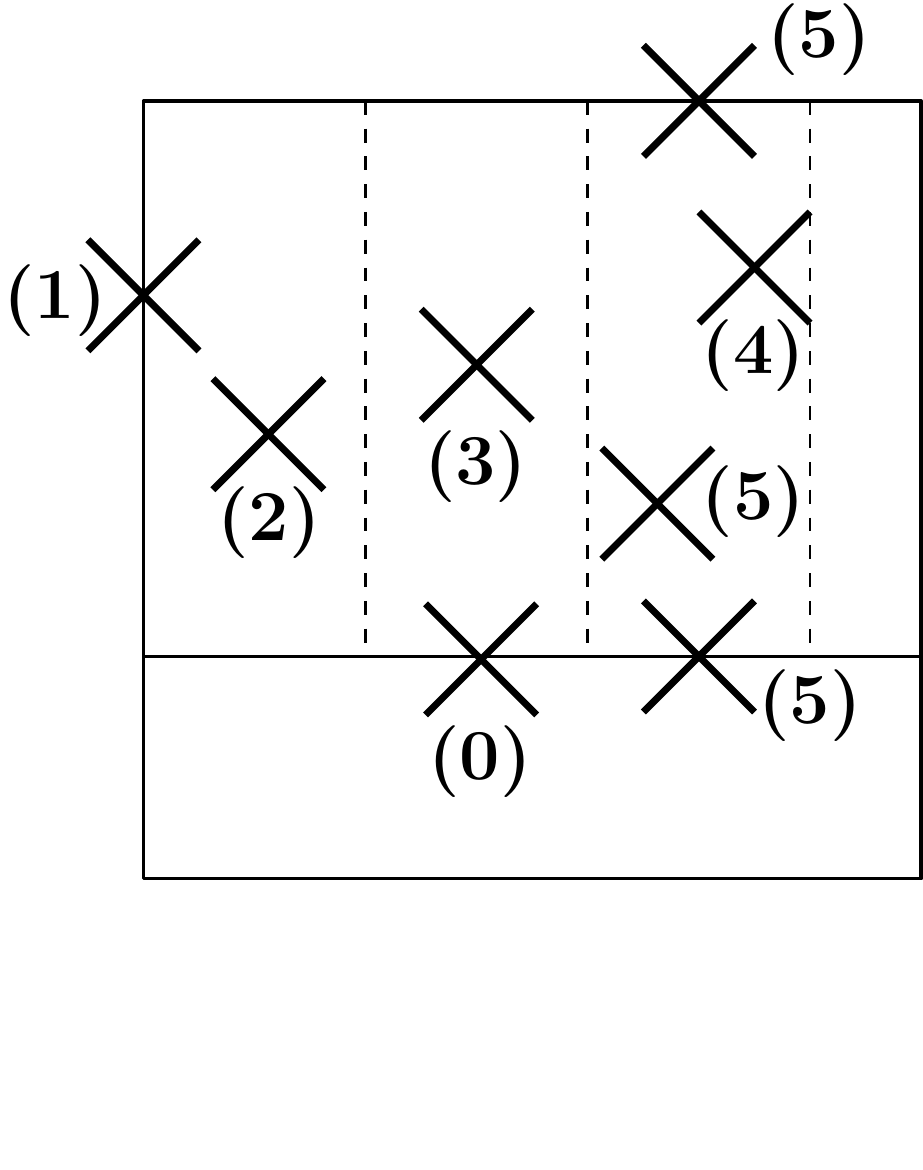} \hspace{2ex}
	\includegraphics[scale=0.31]{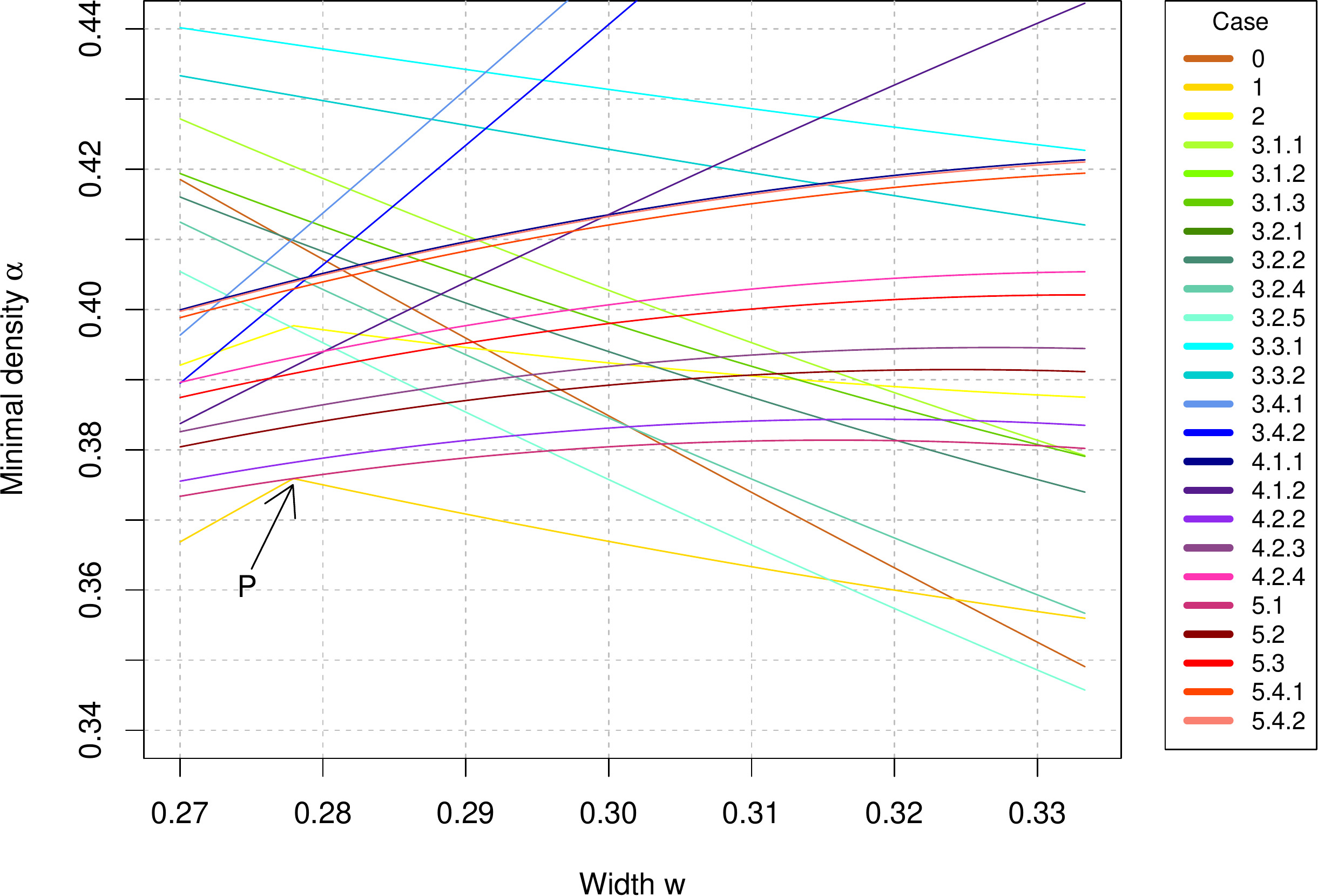}
	\caption[Cases for the analysis of packing into a square]{(Left): Different cases for an overlap. Case $0$: A single circle is too large for $L_0$. Case $1$: $L_0$ exceeded. Case $2$: Overlap in $L_2$. Case $3$: Overlap in $L_3$. Case $4$: Overlap in $L_4$ with large circle being involved. Case $5$: Overlap in $L_4$ with no large circle being involved. (Right): \new{A plot of 24 terms for corresponding 24 (sub-)cases 
	for $\qS=0.191578$}. The point $P=(0.277927,0.375898)$ is the highest point of the $0$-level. Its $y$-value is the highest guaranteed packing density for circles with \new{minimal radii of $0.191578 \cdot 0.277927/2 < 0.0266223$.}}
	\label{fig:cases_overview2}
\end{figure}

In this section, we analyze the packing density of our overall approach for online packing circles into the unit square. In order to prove Theorem~\ref{thm:mainsquare_notiny}, i.e., a lower bound for the achieved packing density, we show that if there is an overlap or if there is no space in the last lane, then the occupied area must be at least this lower bound. Our analysis distinguishes six different \new{higher-level} cases of where and how the overlap can happen, see \hyperref[fig:cases_overview2]{Fig.~\ref*{fig:cases_overview2}} left.

For each of the six cases and its subcases, we explicitly give a density function, providing the guaranteed packing density depending on the choice of $w$, see Fig.~\ref{fig:cases_overview2} right. \new{The shown functions are constructed for the case of no (very) tiny circles with an alternative $q_2=0.191578$, which was chosen numerically in order to find a high provable density. The $w$ with the highest guaranteed packing density of $0.375898$ is $w=0.277927$.
This concludes the proof of Theorem~\ref{thm:mainsquare_notiny}.}

\restatethm{\thmmainsquarenotiny*}{thm:mainsquare_notiny}

\new{With the same idea but with $w=0.288480$ and all circles classes, especially with classes $i\ge 2$ as defined in Table~\ref{table:lane_sizes}, we prove Theorem~\ref*{thm:mainsquare}}.


\restatethm{\thmmainsquare*}{thm:mainsquare}


\section{Conclusion}\label{sec:conclusion}
We provided online algorithms for packing circles into \new{a square and a rectangle}. For
the case of a rectangular container, we guarantee a packing density which is
worst-case optimal for rectangles with a skew of at least $2.36$. For the case
of a square container, we provide a packing density of $0.350389$ which we
improved to $0.375898$ if the radii of incoming circles are lower-bounded
by $0.026622$.

%
%
%
 \bibliography{refs}

\begin{thebibliography}{10}

\bibitem{Brubach14}
B.~Brubach.
\newblock Improved online square-into-square packing.
\newblock In {\em Proc. 12th International Workshop on Approximation and Online
  Algorithms (WAOA)}, pages 47--58, 2014.

\bibitem{Castillo2008}
I.~Castillo, F.~J. Kampas, and J.~D. Pint\'{e}r.
\newblock Solving circle packing problems by global optimization: numerical
  results and industrial applications.
\newblock {\em European J. of Operational Research}, 191(3):786--–802, 2008.

\bibitem{DPW1990optimal}
C.~de~Groot, R.~Peikert, and D.~W{\"u}rtz.
\newblock The optimal packing of ten equal circles in a square.
\newblock Technical Report 90-12, ETH Zurich, Switzerland, 1990.

\bibitem{Fekete10proceeding}
E.~D. Demaine, S.~P. Fekete, and R.~J. Lang.
\newblock Circle packing for origami design is hard.
\newblock In {\em Proceedings 5th International Conference on Origami in
  Science, Mathematics and Education (Origami$^5$)}, pages 609--626. A. K.
  Peters/CRC Press, 2010.

\bibitem{Fekete11}
S.~P. Fekete and H.~Hoffmann.
\newblock Online square-into-square packing.
\newblock {\em Algorithmica}, 77(3):867--901, 2017.

\bibitem{fekete:circlesintocircles}
S.~P. Fekete, P.~Keldenich, and C.~Scheffer.
\newblock Packing disks into disks with optimal worst-case density.
\newblock In {\em Proc. 35th Symposium on Computational Geometry (SoCG)}, 2019.
\newblock to appear.

\bibitem{Fekete18}
S.~P. Fekete, S.~Morr, and C.~Scheffer.
\newblock Split packing: Algorithms for packing circles with optimal worst-case
  density.
\newblock {\em Discrete \& Computational Geometry}, 2018.
\newblock Online First at https://doi.org/10.1007/s00454-018-0020-2.

\bibitem{Fraser1994}
H.~J. Fraser and J.~A. George.
\newblock Integrated container loading software for pulp and paper industry.
\newblock {\em European J. of Operational Research}, 77(3):466--474, 1994.

\bibitem{George95}
J.~A. George, J.~M. George, and B.~W. Lamar.
\newblock Packing different-sized circles into a rectangular container.
\newblock {\em European J. of Operational Research}, 84(3):693--712, 1995.

\bibitem{Leung90}
J.~Y.~T. Leung, T.~W. Tam, C.~S. Wong, G.~H. Young, and F.~Y. Chin.
\newblock Packing squares into a square.
\newblock {\em Journal of Parallel and Distributed Computing}, 10(3):271--275,
  1990.

\bibitem{LR2002packing}
M.~Locatelli and U.~Raber.
\newblock Packing equal circles in a square: a deterministic global
  optimization approach.
\newblock {\em Discrete Applied Mathematics}, 122(1):139--166, 2002.

\bibitem{MC2005new}
M.~C. Mark{\'o}t and T.~Csendes.
\newblock A new verified optimization technique for the" packing circles in a
  unit square" problems.
\newblock {\em SIAM J. on Optimization}, 16(1):193--219, 2005.

\bibitem{Moser67}
J.~W. Moon and L.~Moser.
\newblock Some packing and covering theorems.
\newblock {\em Colloquium Mathematicae}, 17(1):103--110, 1967.

\bibitem{NO1998more}
J.~K. Nurmela and J.~P.~R. Osterg{\aa}rd.
\newblock More optimal packings of equal circles in a square.
\newblock {\em Discrete \& Computational Geometry}, 22(3):439--457, 1998.

\bibitem{oler}
N.~Oler.
\newblock A finite packing problem.
\newblock {\em Canad. Mathematical Bulletin}, 4:153--–155, 1961.

\bibitem{schaer1965densest}
J.~Schaer.
\newblock The densest packing of nine circles in a square.
\newblock {\em Canad. Mathematical Bulletin}, 8:273--277, 1965.

\bibitem{SM1965geometric}
J.~Schaer and A.~Meir.
\newblock On a geometric extremum problem.
\newblock {\em Canad. Mathematical Bulletin}, 8(1), 1965.

\bibitem{Sugihara2004}
K.~Sugihara, M.~Sawai, H.~Sano, D.~S. Kim, and D.~Kim.
\newblock Disk packing for the estimation of the size of a wire bundle.
\newblock {\em Japan Journal of Industrial and Applied Mathematics},
  21(3):259--278, 2004.

\bibitem{Szabo07}
P.~G. Szab{\'o}, M.~C. Mark{\'o}t, T.~Csendes, E.~Specht, L.~G. Casado, and
  I.~Garc{\'\i}a.
\newblock {\em New Approaches to Circle Packing in a Square: With Program
  Codes}, volume~6 of {\em Springer Optimization and Its Applications}.
\newblock Springer, 2007.

\bibitem{Peikert1994}
D.~W\"{u}rtz, M.~Monagan, and R.~Peikert.
\newblock The history of packing circles in a square.
\newblock {\em Maple Technical Newsletter}, pages 35--42, 1994.

\end{thebibliography}

\end{document}